\documentclass[twocolumn,preprintnumbers,amsmath,amssymb,superscriptaddress]{revtex4}

\usepackage{graphicx}
\usepackage{dcolumn}
\usepackage{bm}
\usepackage{color}
\begin{document}

\title{Observation of coherent oscillation in single-passage Landau-Zener
transitions}
\author{Guozhu Sun}
\email{gzsun@nju.edu.cn}
\affiliation{National Laboratory of Solid State Microstructures and Research Institute of Superconductor Electronics,School of Electronic Science and Engineering, Nanjing University, Nanjing 210093, China}
\affiliation{Synergetic Innovation Center of Quantum Information and Quantum Physics, University of Science and Technology of China, Hefei, Anhui 230026, China}
\affiliation{Department of Physics and Astronomy, University of Kansas, Lawrence, KS 66045, USA}

\author{Xueda Wen}
\affiliation{Department of Physics, University of Illinois at Urbana-Champaign, Urbana, IL 61801, USA}
\author{Ming Gong}
\affiliation{Department of Physics and Astronomy, University of Kansas, Lawrence, KS 66045, USA}
\affiliation{National Laboratory of Solid State Microstructures, School of Physics, Nanjing University, Nanjing 210093, China}
\author{Dan-Wei Zhang}
\affiliation{National Laboratory of Solid State Microstructures, School of Physics, Nanjing University, Nanjing 210093, China}
\author{Yang Yu}
\affiliation{National Laboratory of Solid State Microstructures, School of Physics, Nanjing University, Nanjing 210093, China}
\affiliation{Synergetic Innovation Center of Quantum Information and Quantum Physics, University of Science and Technology of China, Hefei, Anhui 230026, China}
\author{Shi-Liang Zhu}
\affiliation{National Laboratory of Solid State Microstructures, School of Physics, Nanjing University, Nanjing 210093, China}
\author{Jian Chen}
\affiliation{National Laboratory of Solid State Microstructures and Research Institute of Superconductor Electronics,School of Electronic Science and Engineering, Nanjing University, Nanjing 210093, China}
\author{Peiheng Wu}
\affiliation{National Laboratory of Solid State Microstructures and Research Institute of Superconductor Electronics,School of Electronic Science and Engineering, Nanjing University, Nanjing 210093, China}
\affiliation{Synergetic Innovation Center of Quantum Information and Quantum Physics, University of Science and Technology of China, Hefei, Anhui 230026, China}

\author{Siyuan Han}
\affiliation{Department of Physics and Astronomy, University of Kansas, Lawrence, KS
66045, USA}

\begin{abstract}

Landau-Zener transition (LZT) has been explored in a variety of physical
systems for coherent population transfer between different quantum states. In recent
years, there have been various proposals for applying LZT to quantum
information processing because when compared to the methods using ac pulse
for coherent population transfer, protocols based on LZT are less
sensitive to timing errors. However, the effect of finite range of qubit
energy available to LZT based state control operations has not been
thoroughly examined. In this work, we show that using the well-known
Landau-Zener formula in the vicinity of an avoided energy-level crossing
will cause considerable errors due to coherent oscillation
of the transition probability in a single-passage LZT experiment. The data
agree well with the numerical simulations which take the transient dynamics of
LZT into account. These results not only provide a closer view on the issue
of finite-time LZT but also shed light on its effects on the quantum state
manipulation.

\end{abstract}

\maketitle

Landau-Zener transition (LZT) has broad applications in atomic and molecular physics, quantum optics,
condensed matter physics, chemical physics, and quantum information science.
For example, LZT has been applied to investigating the jump time and quantum
Zeno and anti-Zeno effects of cold atoms in accelerated optical lattices
\cite{PhysRevLett.87.040402,PhysRevLett.103.090403}, the behavior of
molecular magnets at low temperature \cite{Wernsdorfer02041999,
PhysRevB.79.104423}, nonequilibrium phase transitions \cite%
{PhysRevLett.95.035701}, and it is also exploited as a tunable beam splitter of wave functions
to generate entangled multipartite states \cite%
{Nat.Commun.,PhysRevLett.110.173603}. LZT also plays a key role in
determining whether random optimization problems can be solved using the
quantum adiabatic algorithm \cite{Bapst2013127}. Recently, LZT's potential
for robust manipulation of coherent quantum states has attracted much
attention in the context of quantum information processing \cite%
{Nature.455.51,PhysRevLett.100.113601,PhysRevLett.103.220401,Shevchenko2010,Nat.Commun.,PhysRevLett.107.207002,J. Phys. A: Math. Theor. 44 4153022,PhysRevLett.109.237005,PhysRevLett.110.173603,NaturePhysics.8.147,New J. Phys. 14 013024,New J. Phys. 16 123024} because LZT may provide a simple and effective solution to the realization of high fidelity quantum state control without the need for precise
timing.

The time-dependent Hamiltonian describing LZT in quantum two-level systems can be written in the generic form as \begin{equation}
H_{LZ}(t)=-\frac{1}{2}\varepsilon (t)\sigma _{z}-\frac{1}{2}\Delta \sigma
_{x},  \label{H-LZT}
\end{equation}%
where $\sigma _{x,z}$ are Pauli matrices, $%
\varepsilon (t)=vt$ is the energy difference between the two diabatic
(crossing) basis states (i.e., the eigenstates $%
|\uparrow\rangle $ and $|\downarrow\rangle $ of the $\sigma _{z}$ operator) controlled by an external
parameter which depends linearly on time $t,$ and $\Delta $ is the constant
gap between the two instantaneous eigenenergy states $|+\rangle $ and $%
|-\rangle $ at the center of the avoided crossing $\varepsilon =0$, as
depicted in Fig. 1(a). In such systems, when $\varepsilon (t)$ is swept
through the avoided crossing, transitions between $|\pm \rangle $ with
energies $E_{\pm }(t)=\pm \frac{1}{2}\sqrt{\varepsilon (t)^{2}+\Delta ^{2}}$
can occur and the transition probability is given by the well-known Landau-Zener
(LZ) formula
\begin{equation}
P_{LZ}=e^{-\frac{\pi \Delta ^{2}}{2v}},  \label{LZ}
\end{equation}%
where $v=|d\varepsilon /dt|$ is the Landau-Zener speed and we have set the
reduced Planck constant $\hbar =1$. Equation (\ref{LZ}) gives the probability
of finding the system in the excited (ground) state at $\varepsilon
_{f}=\varepsilon (t\rightarrow +\infty )$ when it is started in the ground
(excited) state at $\varepsilon _{i}=\varepsilon (t\rightarrow -\infty ).$
By defining $\alpha =\Delta ^{2}/4v$\ as the adiabaticity parameter the LZ
formula can be simplified to $P_{LZ}=\exp (-2\pi \alpha )$. Although
analytical solution to the problem cannot be obtained when $\varepsilon _{i}$
and/or $\varepsilon _{f}$ are finite, it is well known that for $%
|\varepsilon _{i,f}|\gg \Delta $, the LZ
formula provides an excellent approximation to the actual
transition probability and $P_\downarrow \approx P_{LZ}$. However, when $|\varepsilon _{i,f}|\gg \Delta $ is not satisfied, the LZ formula may become quantitatively
inaccurate or even qualitatively incorrect. In spite of some theoretical
studies on the effects of finite $|\varepsilon _{i,f}|/\Delta $ on $P_\downarrow$, there is an acute lack of adequate experimental evidence.

On the other hand, understanding LZT with moderate values of $|\varepsilon _{i,f}|/\Delta $ is in
urgent need because this region of parameter space is important to quantum
information processing. For instance, in superconducting qubits the tuning
range of energy level spacing is usually limited to a couple of GHz or even
as narrow as a few hundreds of MHz while $\Delta /2\pi $ could be as large
as $10^{2}$ MHz \cite%
{WilliamD.Oliver12092005,PhysRevLett.96.187002,PhysRevLett.98.257003,PhysRevLett.101.017003,PhysRevLett.101.190502,sun:102502,Nature.459.960,Nat.Commun.,PhysRevB.83.180507,PhysRevLett.110.173603}%
. For quantum state control based on sweeping $\varepsilon$ through avoided
crossings, understanding LZT probability's dependence on $|\varepsilon
_{i,f}|/\Delta $ and the sweeping time is essential to high fidelity operation. The fidelity of
various techniques based on LZT relies critically on the accuracy of the LZ
formula which predicts a simple exponential dependence of $P_{LZ}$ on the
adiabaticity parameter $\alpha $ only. Therefore, for a qubit starting from
the ground (excited) state the probability of finding it in the excited
(ground) state after a \textit{single passage} through the avoided crossing
is assumed to be determined entirely by $\alpha $ (i.e., $\Delta ^{2}/4v$) but not the detail of
the process such as $|\varepsilon _{i,f}|/\Delta $. Here,
using an artificial atom --- a superconducting phase qubit --- coupled to a
microscopic two-level system (TLS), we test the accuracy of the LZ formula in the region of $|\varepsilon _{i,f}|/\Delta <4.3$. We show that
in contrast to conventional wisdom, in the region of parameter space most
relevant to superconducting qubits, $P_\downarrow$ could deviate significantly
from $P_{LZ}$ determined by the LZ formula. Our experiment and numerical simulation demonstrate $P_\downarrow$ can oscillate coherently as a function of $\varepsilon _{f}$ for
constant $\alpha$ when $|\varepsilon _{i,f}|$ is comparable to $\Delta$, which is named as coherent Landau-Zener oscillation (LZO).

\noindent\textbf{Results}

In our experiment we use a superconducting phase qubit. However, since a
single phase qubit does not have an intrinsic avoided energy-level crossing,
we utilize an avoided level crossing arising from interaction between the qubit
and a microscopic TLS \cite{QuantumInfProcess.8.81,QuantumInfProcess.8.117}. As discussed below in more detail, when the transition
frequency of the qubit $\omega _{10}$ is close to that of the TLS $\omega
_{TLS},$ which is fixed, the first and second excited states of the coupled
qubit-TLS system form an effective quantum two-level system described by the
LZ Hamiltonian (\ref{H-LZT}). Note that to make quantitative
comparisons between the theory/numerical simulation and the experiment without free
parameters, all relevant system parameters, including the energy relaxation
and dephasing time of the qubit and the energy gap $\Delta ,$ are obtained from
direct measurements.

A microscopic picture of the superconducting phase qubit is shown in Fig. 1(b). The qubit
 consists of an $L\approx 770$ pH superconducting loop intersected by a
Al/AlO$_{x}$/Al Josephson tunnel junction with a critical current $%
I_{0}\approx 1.4$ $\mu $A and a junction capacitance $C\approx 240$ fF. By
varying the magnetic flux applied to the superconducting loop the potential
energy of the qubit becomes asymmetrical. The ground state and the first
excited state in the upper potential well, represented by $|0\rangle $ and $%
|1\rangle $ respectively, can be used as the computational basis states of
the qubit. For an isolated qubit, the transition frequency
between $|0\rangle $ and $|1\rangle ,$ $\omega _{10},$ is a single-valued
function of the external flux bias $\Phi _{x}$ which is coupled inductively
to the superconducting loop through an on-chip flux bias line.

As shown in Fig. 2(a), however, the microwave spectrum of the qubit $%
\omega _{10}(\Phi _{x})$ has a rather large avoided energy-level crossing at
$\Phi _{x}\approx -2.8$ m$\Phi _{0}$ (with respect to the flux bias point at
which $\omega _{10}/2\pi\approx 16.348$ GHz) indicating significant interaction
between the qubit and a microscopic TLS \cite{PhysRevB.82.132501}. The
transition frequency between the TLS' ground state $|g\rangle $ and excited
state $|e\rangle $ and the qubit-TLS coupling strength are $%
\omega _{TLS}/2\pi =16.450\pm 0.002$ GHz and $\Delta /2\pi =70.0\pm 0.5$ MHz
from the spectrum and vacuum Rabi oscillation, respectively. Note that in
this coupled qubit-TLS system the time-dependent energy difference between
the two diabatic states involved in LZT is $\varepsilon (t)=\omega
_{10}(t)-\omega _{TLS}$ which depends linearly on the flux bias to a good
approximation. The relationship between the flux bias and $\varepsilon $ can
be found from Fig. 2(a).


Fig. 1(c) illustrates the experimental procedure used to observe coherent LZO. We
begin by setting the initial diabatic energy of the effective quantum
two-level system $\varepsilon _{i}$ at about $100$ MHz below $\omega _{TLS}$
with a static flux bias. \ The qubit is prepared in its ground state by
waiting for much longer than the energy relaxation time $%
T_{1}\approx 70$ ns of the qubit. A microwave pulse is then applied
to the qubit when it is biased at a fixed value $\varepsilon _{i}/2\pi\approx
-100 $ MHz. The microwave pulse coherently transfers the population of the
qubit-TLS from $|0g\rangle $ to one of the system's eigenstates $|-\rangle $ through
a process that is discussed in detail in Methods. The lack of oscillation in $T_{1}$ measurement taken at $%
\varepsilon _{i}$ as shown in Fig. 2(b) confirms that the initial state of the
qubit-TLS system at $t=0$ is indeed the eigenstate $|-\rangle .$
As illustrated in Fig. 1(c), a time-dependent flux $\Phi (t)=\Phi
_{LZ}t/t_{sp}$ is then superimposed between $t=0$ and $t=t_{sp}$ onto the
static flux bias to sweep $\varepsilon $ linearly from $\sim
-100$ MHz to its maximum value $\varepsilon _{f}$. The corresponding LZ
speed $v$ is thus $(\varepsilon _{f}-\varepsilon _{i})/t_{sp}$. This is followed
immediately by a $5$-ns readout pulse which performs a projective
measurement of the probability $P_\downarrow$ of finding the qubit in state $%
|1\rangle $ (i.e., the coupled system is in state $|1g\rangle$ corresponding to $|\downarrow\rangle$ in Fig. 1(a)).


We first measure $P_\downarrow$ vs. $t_{sp}$ at a constant value
of $\varepsilon _{f}$ by keeping $\Phi _{LZ}$ fixed while increasing $t_{sp}$
from almost $0$ ns to $45$ ns. The maximum $t_{sp}$ is selected to avoid too much
influence of the qubit's energy relaxation. By stepping $\Phi _{LZ}$ from $0$
to $-11$ m$\Phi _{0}$ the value of corresponding $\varepsilon _{f}$ is then
varied from about $-1.4\Delta $ to $\sim 4\Delta $ $,$ in which the
condition $|\varepsilon _{f}|\gg \Delta $ is no longer satisfied.
This procedure is repeated at each $\varepsilon _{f}$ to obtain $P_\downarrow(\varepsilon _{f},t_{sp})$.
Fig. 3(a) shows the dependence of $P_\downarrow$ on $\varepsilon _{f}$ and $t_{sp}$.
It can be seen that $P_\downarrow$ vs. $%
t_{sp}$ decays exponentially for $\varepsilon _{f}<0$ ($\Phi _{LZ}\in
\lbrack -2.8$ m$\Phi _{0},$ $0]$) with a characteristic
time $T_{1}$ due to energy relaxation. As $\varepsilon _{f}$ becomes
positive, $P_\downarrow$ vs. $t_{sp}$
becomes oscillatory. Since the avoided crossing is traversed only once, the
observed oscillation in $P_\downarrow$ vs. $t_{sp}$ with constant $\varepsilon _{f}$
must be a consequence of the
moderate value of $\varepsilon _{f}/\Delta $ and is not caused by the Landau-Zener-St\"{u}ckelberg interference which requires multiple
passages through the avoided crossing.

It is worth noting that the observed oscillation is not a
consequence of ill-prepared initial states with non-negligible probability
amplitude in the excited state $|+\rangle $ of the effective
Hamiltonian (\ref{H-LZT}) because the microwave pulse used to initialize the
system resonantly couples $|0g\rangle $ to $|-\rangle $ and has negligible
coupling to $|+\rangle $ due to large frequency detuning. Furthermore,
this process of transferring the system to the desired initial state $|-\rangle
$ via a resonant microwave pulse is robust in the sense that it does not
depend sensitively on the accuracy of the pulse duration $t_{MW}.$
Deviation in pulse duration simply leaves some probability amplitude in $%
|0g\rangle $ which has no effect on LZT other than reducing the visibility
of the oscillation (see Methods for detail on the initial state preparation).
Therefore, we are confident that the oscillation observed in the
non-adiabatic region of the parameter space arises neither from
Landau-Zener-St\"{u}ckelberg interference nor unwanted probability amplitude
of $|+\rangle $ in the initial state. This is also supported by the good agreement between the results of experiment and numerical simulation shown in Fig. 3(b), which uses $|-\rangle $ as the initial state at the start of the single passage sweep.

\noindent\textbf{Discussion}

By replacing $\varepsilon _{f}$ with $\varepsilon _{i}+vt$, solving the
problem of sweeping $\varepsilon $ in a finite range is transformed to
finding $P_\downarrow$ at finite time. Previous studies have discovered that LZ
transition probability reaches the asymptotic value given by equation (\ref{LZ})
at $t\gg t_{LZ}$ (assume $\varepsilon =0$ at $t=0$)\textrm{, }where $t_{LZ}=2%
\sqrt{\alpha }/\Delta \max (1,\sqrt{\alpha })$ is called the Landau-Zener
time, and oscillates in the vicinity of avoided crossing (corresponds to $t \leq t_{LZ})$ due to the transient dynamics \cite{PhysRevA.23.3107,PhysRevLett.62.2543,Shevchenko2010}. Since corrections to the standard LZ formula are significant only if the adiabaticity parameter $\alpha \leq 1$, the region of $\varepsilon $ within which transient dynamics
plays an important role is given by $\varepsilon _{f} \leq t_{sp}^{-1},$
where ``$\leq $'' means less than or comparable to. By examining the
experimental data we find that the region of most noticeable coherent LZO coincides with $\varepsilon _{f}\leq t_{sp}^{-1}$ which agrees
well with the result of numerical simulation. These results unambiguously show that coherent LZO is originated from the transient dynamics of the
LZT.

Such coherent LZO has little effect on the adiabatic evolution.
Because in the true adiabatic regime, by definition the system always stays in the
instantaneous ground state and no LZT could occur. In order to find the
region of approximate adiabatic evolution in our experiment, it is necessary to modify the
definition of the adiabaticity parameter to $\alpha' =\frac{(\omega
_{TLS}-\omega _{10})^{2}+\Delta ^{2}}{4v}$. For the adiabatic
theorem to hold $\alpha' \gg 1$ is required. As shown in Fig. 3(a),
the white dashed lines represent $\alpha' =10$. It is
clear that there is no coherent LZO observed in the region $\alpha'
\gg 1$.

In a popular analogy to optics an avoided crossing acts as an effective
beam splitter, with a transmission coefficient corresponding to $P_{LZ}$ in the LZ formula, for quantum wave functions. This beam splitter analogy
has been applied successfully to the visualization and explanation of the behavior
of superconducting and semiconductor qubits \cite%
{EPL.65.844,WilliamD.Oliver12092005,PhysRevLett.96.187002,Nature.455.51,
Nat.Commun.,J.R.Petta02052010,NaturePhysics.8.54,Nat.Commun.Guo}. In this analogy, a single sweep
through the avoided crossing is equivalent to passing a beam of light
through the beam splitter only once. When $%
|\varepsilon _{i,f}|\gg \Delta $, $P_\downarrow \approx P_{LZ}$ and thus a greater LZ speed corresponds
to a higher transmission coefficient of the beam splitter according
to the LZ formula. But when $|\varepsilon _{i,f}|\gg \Delta $ is not satisfied, $P_\downarrow$ differs greatly from $P_{LZ}$. As an example, $P_\downarrow$ vs. $t_{sp}$, and thus the LZ speed $v$, with $\varepsilon _{f}/2\pi=200$ MHz is shown in Fig. 4(a). The maximum in the difference $\delta P_\downarrow$ between the experimental $P_\downarrow$ and those obtained from the LZ formula (2), shown in the inset of Fig. 4(a), can reach 0.21. The observation of coherent LZO strongly
suggests that when $|\varepsilon _{i,f}|\gg \Delta $ is not met corrections to the LZ formula should be considered to avoid conceptual
difficulties.

Coherent LZO also has significant consequences on the coherent
manipulation of quantum states of single qubits and coupled two-qubit systems based on LZT \cite{PhysRevLett.100.113601,Shevchenko2010}. For
this approach of quantum state control, the LZ transition probability $P_{LZ}$
plays a central role since each single passage through the avoided crossing
results in a unitary operation $U_{LZ}$ given by \cite%
{Shevchenko2010}
\[
U_{LZ}=\left(
\begin{array}{cc}
i\sqrt{1-P_{LZ}}e^{-i\varphi _{s}} & -\sqrt{P_{LZ}} \\
\sqrt{P_{LZ}} & -i\sqrt{1-P_{LZ}}e^{i\varphi _{s}}%
\end{array}%
\right) ,
\]%
where $\varphi _{s}$ is the Stokes phase \cite{PhysRevA.55.R2495}, which has
no effect on the single-passage LZT process discussed here and thus can be set to zero for the sake of convenience. As mentioned above,
the transition frequency $\omega _{10}$ of most artificial atoms, in
particular the superconducting qubits, is limited to a couple of GHz. Because
the speed of two-qubit operations is proportional to the inter-qubit coupling
strength $\Delta $, increasing $|\varepsilon _{i,f}|/\Delta $ by reducing $%
\Delta $ is undesirable. Hence, evaluating $U_{LZ}$ according to the LZ
formula (\ref{LZ}) could result in significant errors when $\varepsilon
_{f}\gg \Delta $ is not satisfied. In order to conduct a quantitative analysis,
the experimental data along the yellow dashed line corresponding to constant $v$, in the Fig. 3(a)
are extracted and shown in Fig. 4(b). Oscillation in $P_\downarrow$ is clearly observed
and it is qualitatively different from the exponential decay predicted by equation (\ref{LZ}), when decoherence is taken into account. Suppose the initial quantum
state is $|\downarrow\rangle$. Then after a single passage through the avoided crossing,
if one replaces $P_{LZ}$ with $P_\downarrow$ as in the asymptotic situation,
the deviation $\delta P_{\downarrow }=$ $P_{\downarrow }-P_{LZ}$ would be quite large. For example, when $%
t_{sp}=12.8$ ns the deviation $\delta P_{\downarrow }=0.229$, which is unacceptably large for coherent quantum state transformation.

In conclusion, we have investigated the effect of finite energy ($\varepsilon $) sweep (or
equivalently finite time) on LZT probability $P_\downarrow$
experimentally. Single-passage technique is used to isolate the effect of
finite $\varepsilon_f$ on $P_\downarrow$ from that of interference caused by
passing the avoided crossing multiple times. We find that $%
P_\downarrow(\varepsilon _{f}/\Delta ,\alpha =const)$ oscillates when $\varepsilon _{f}$ is comparable
to $\Delta $ and $\alpha <1$. The good agreement between the experiment and numerical calculation strongly supports the notion that coherent LZO is caused by the underlying
transient dynamics of the finite time LZT which cannot be described by the LZ asymptotic formula. In this region of the LZT parameter space, corrections to the LZ formula must be taken into account, otherwise it will lead to substantial errors in quantum state operations based on LZT. The result also shows that when applying the simple beam splitter analogy one should not
automatically assume that greater $\alpha $ (i.e., faster sweep) corresponds
to larger transmission coefficient (i.e., greater $P_{LZ}$) as implied by the asymptotic LZ formula.

\noindent\textbf{Methods}

\noindent\textbf{Initial state preparation}

We first derive an analytical result explaining the lack of
oscillation at the very beginning of LZT. The Hamiltonian of the qubit-TLS system coupled to a microwave field is given as: (in the basis $\{|0g\rangle, |1g\rangle, |0e\rangle,
|1e\rangle\}$)
\begin{equation}
H=
\left(\begin{array}{cccc}
0       &\Omega_m\cos\omega t       &0      &0\\
\Omega_m\cos\omega t &\omega_{10}       &\Delta/2      &0\\
0 &\Delta/2          &\omega_{TLS}    &\Omega_m\cos\omega t\\
0       &0      &\Omega_m\cos\omega t            & \omega_{10}+\omega_{TLS} \\
\end{array} \right),
\label{H1}
\end{equation}
where $\Omega_m$ is the Rabi frequency, $\omega$ is the microwave frequency, $\omega_{TLS}$ is the energy difference between the ground state $|g\rangle$ and the excited state $|e\rangle$ of TLS, $\Delta$ is the coupling strength between the qubit and TLS,
$\delta=\omega_{10}-\omega$ and $\delta_r=\omega_{TLS}-\omega_{10}$ are detunings.
By rotating the frame, Hamiltonian (\ref{H1}) can be transformed to the following time-independent form \cite{PhysRevB.80.094507,PhysRevB.82.132501}
\begin{equation}
H_1=\frac{1}{2}
\left(\begin{array}{cccc}
-\delta_r-2\delta      &\Omega_m     &0      &0\\
\Omega_m  &-\delta_r     &\Delta      &0\\
0 &\Delta          &\delta_r    &\Omega_m\\
0       &0      &\Omega_m        & \delta_r+2\delta  \\
\end{array} \right).
\end{equation}
Next, we rewrite $H_1$ in which the subspace spanned by $\{|1g\rangle, |0e\rangle\}$ is diagonalized:
\begin{equation}\small
H_2=
\left(\begin{array}{cccc}
-\frac{\delta_r+2\delta}{2}     &-\frac{\Delta}{4N_-}\Omega_m    &-\frac{\Delta}{4N_+}\Omega_m       &0\\
-\frac{\Delta}{4N_-}\Omega_m   &-\frac{1}{2}\sqrt{\Delta^2+\delta_r^2}     &0     &-\frac{\Delta}{4N_+}\Omega_m \\
-\frac{\Delta}{4N_+}\Omega_m  &0         &\frac{1}{2}\sqrt{\Delta^2+\delta_r^2}    &-\frac{\Delta}{4N_-}\Omega_m \\
0       &-\frac{\Delta}{4N_+}\Omega_m       &-\frac{\Delta}{4N_-}\Omega_m        & \frac{\delta_r+2\delta}{2}   \\
\end{array} \right),
\end{equation}
where $N_{\mp}=\sqrt{\Delta^2/4+(-\delta_r/2\pm \sqrt{\Delta^2+\delta_r^2}/2)^2}$.
The basis for $H_2$ is now $\{|0g\rangle, |-\rangle, |+\rangle, |1e\rangle\}$.
Note that in our
experiment, before turning on the microwave the state is at $\Psi (t=0)=|0g\rangle $. By turning on the microwave ($\Omega_m\neq 0$), $|0g\rangle$ is coupled to both $|-\rangle$ and $|+\rangle$. The resonance between $|0g\rangle$ and $|\pm\rangle$ occurs when
$-\frac{\delta_r+2\delta}{2}=\pm \sqrt{\Delta^2+\delta_r^2}/2 $, from which we obtain the resonant condition:
\begin{equation}
\omega=\frac{\omega_{10}+\omega_{TLS}}{2}\pm \frac{1}{2}\sqrt{(\omega_{TLS}-\omega_{10})^2+\Delta^2} \equiv \lambda_{\pm}.
\end{equation}

In the limit of $\omega_{10}-\omega_{TLS}\gg \Delta$, we have $\omega=\omega_{10}$, which corresponds to the usual two-state Rabi oscillation. Note that in this limit there is also a solution $\omega=\omega_{TLS}$. However, in this case the coupling strength is $-\frac{\Delta}{2N_+}\Omega_m \to 0$. The reason is that
although the microwave frequency could match that of TLS, coupling between the microwave and TLS is negligible which is confirmed by the absence of
Rabi oscillation between the two states of the TLS in a separate experiment.
In the other limit of $\omega _{10}-\omega _{TLS}=0$,
we have $\omega =\frac{\omega _{10}+\omega _{TLS}}{2}\pm \Delta $,
and the dynamics have been thoroughly studied in Ref. \cite{PhysRevB.82.132501}.

In our experiment, we have $(\omega_{TLS}-\omega_{10})/2\pi\approx 100$ MHz and $\Delta/2\pi\approx $ 70 MHz, which means LZT occurs in the region where $(\omega_{TLS}-\omega_{10}) \sim \Delta$. Because the frequency of the applied microwave is $\omega=\lambda_-$, which can be determined from the measured energy spectrum shown in Fig. 2(a), $|0g\rangle$ is resonantly coupled to $|-\rangle$, which is the eigenstate of $H_b=\omega_{10}|1g\rangle\langle 1g|+\omega_{TLS}|0e\rangle\langle 0e|+\frac{\Delta}{2}(|1g\rangle\langle 0e|+|0e\rangle\langle 1g|)$.
 Although there is in principle also a
coupling between $|0g\rangle $ and $|+\rangle $, the effective coupling is
much smaller because of the large detuning, as discussed below.

For $\Delta /2\pi \approx 70$ MHz, $(\omega _{TLS}-\omega
_{10})/2\pi \approx 100$ MHz, the resonance between $|0g\rangle $
and $|-\rangle $ occurs at $\delta /2\pi \simeq 11$ MHz, we obtain $N_{-}/2\pi \simeq 36.7$ MHz and
$N_{+}/2\pi \simeq 116.4$ MHz. In our experiments, the coupling strength
between $|0g\rangle $ and $|-\rangle $ is about $20$ MHz and that between $|0g\rangle $ and $|+\rangle $%
would be $\frac{N_{-}}{N_{+}}\times 20$ MHz $\simeq 6.3$ MHz. Because $6.3$ MHz is comparable with $20$ MHz,
one may think that coupling between $|0g\rangle $ and $|+\rangle $ cannot be neglected.
However, there is also a large
detuning of about $122$ MHz between $|0g\rangle $ and $|+\rangle $.
Therefore, the effective coupling between $|0g\rangle $ and $|+\rangle $ is reduced to $(6.3^{2}/122)\approx 0.33$ MHz and thus can be safely neglected. To be more precise, we
calculated the population $P_{\pm }$ (where $P_{\pm }$ is the population of state $|\pm \rangle $)
after the application of a $\pi $ pulse numerically, and it is found that $P_{+}/P_{-}\simeq 5\times 10^{-5}$. Based on this analysis, when $\omega=\lambda_-$, the dynamics can be described by the Hamiltonian in the subspace $\{|0g\rangle, |-\rangle\}$:
\begin{equation}
H_3=
\left(\begin{array}{cccc}
-\frac{\delta_r+2\delta}{2}     &-\frac{\Delta}{4N_-}\Omega_m    \\
-\frac{\Delta}{4N_-}\Omega_m   &-\frac{1}{2}\sqrt{\Delta^2+\delta_r^2}    \\
\end{array} \right).
\end{equation}
At the resonance $-\frac{\delta_r+2\delta}{2}=-\frac{1}{2}\sqrt{\Delta^2+\delta_r^2} $, $H_3$ becomes
\begin{equation}
H_3=-\frac{1}{2}\sqrt{\Delta^2+\delta_r^2} \times I
+
\left(\begin{array}{cccc}
0    &-\frac{\Delta}{4N_-}\Omega_m    \\
-\frac{\Delta}{4N_-}\Omega_m   &0   \\
\end{array} \right),
\end{equation}
where $I$ is a $2\times 2$ identity matrix.

For initial state $\Psi(t=0)=|0g\rangle$, the amplitude of $|-\rangle$ is
\begin{equation}
C_{-}=i\sin \frac{\Delta}{4N_-}\Omega_m t.
\end{equation}
Considering $|-\rangle=-\frac{\Delta}{2N_-}|1g\rangle+\frac{-\delta_r+ \sqrt{\Delta^2+\delta_r^2}}{2N_-}|0e\rangle$, the amplitude of $|1\rangle$ is
\begin{equation}
C_1=C_{1g}+C_{1e}\approx C_{1g}=-i\frac{\Delta}{2N_-}\sin \frac{\Delta}{4N_-}\Omega_m t. \nonumber\\
\end{equation}
Therefore, with a microwave pulse of duration $t_{MW}$ which is used to prepare the initial state, we have
\begin{equation}
P_{\downarrow}=|C_1|^2 \approx \frac{\Delta^2}{4N_-^2}\left( \sin \frac{\Delta}{4N_-}\Omega_m t_{MW}\right)^2.
\label{P1}
\end{equation}

This is the reason why in the experiment we observe a usual Rabi oscillation instead of Rabi beating \cite{PhysRevB.82.132501} which is indicated by the red circles, as shown in Fig.2(b). In addition, when microwave is turned off, the subspace $\{|+\rangle, |-\rangle\}$ is isolated from $|0g\rangle$ and $|1e\rangle$. Projected into the subspace $\{|+\rangle, |-\rangle\}$, the system stays in the eigenstate $|-\rangle$. This explains why we observe a monotone decay of $P_{\downarrow}$ with no oscillations, as indicated by the blue triangles in Fig.2(b).

\noindent\textbf{Effect of $t_{MW}$ and $\varepsilon_f/\Delta$ on LZT}

In this section, we discuss two factors that may affect the LZT probability, i.e.,
the width of the microwave pulse $t_{MW}$ used to prepare
the initial state at $\varepsilon _{i}$ and the end of the
normalized diabatic energy sweeping $\varepsilon _{f}/\Delta $,
respectively.

After a microwave pulse, by projecting into the subspace $\{|1g\rangle,|0e\rangle\}$, the system is in the eigenstate
$|-\rangle$. Then the dynamics of LZT can be described by $H_b$ with a time-dependent $\omega_{10}(t)$, i.e.,
\begin{equation}
H_b=
\left(\begin{array}{cccc}
\omega_{10}(t)    &\Delta/2   \\
\Delta/2   &\omega_{TLS}   \\
\end{array} \right).
\end{equation}
To investigate the Landau-Zener diffraction effect, we sweep $\omega_{10}(t)$ across $\omega_{TLS}$, i.e.,
\begin{equation}
\omega_{10}(t)=\omega_{10}(t=0)+vt, (0\le t\le t_{sp}).
\end{equation}
When $\omega_{10}(t=t_{sp})-\omega_{TLS}\gg \Delta$, we expect that the Landau-Zener asymptotic formula holds and $P_\downarrow \approx P_{LZ}$. Thus no oscillations should occur in $P_\downarrow$. This is confirmed by the result of
numerical simulation shown in Fig. 5(c), where $\varepsilon _{f}/2\pi=1450$ MHz $\approx 20.7\Delta $,
reproducing the exponential decay behavior described by the asymptotic LZ
formula independent of the initial state of the qubit-TLS system. In this case, the population in the qubit state $|1\rangle$ can be expressed as
\begin{equation}
P_\downarrow\propto \frac{\Delta^2}{4N_-^2}\left( \sin \frac{\Delta}{4N_-}\Omega_m t_{MW}\right)^2\times e^{-\frac{\pi\Delta^2}{2v}}\times e^{-\gamma(t_{MW}+t_{sp})}
\end{equation}
where the first term reflects the effect of microwave duration $t_{MW}$ in preparing the initial state, the second term
corresponds to the LZT probability, and the third term represents the relaxation effect. However, as $\omega _{10}(t=t_{sp})$ moves towards $\omega _{TLS}$, the
situation $\omega _{10}(t=t_{sp})-\omega _{TLS}\gg \Delta $ does not hold
any more, and we observe oscillation features in the $t_{sp}$ direction, as
shown in Fig. 5(a) (experiment) and 5(b) (numerical simulation). Notice that Fig. 5(a)
and 5(b) also confirm that the effect of
imprecise $\pi $ pulse is an incomplete transfer of system from $|0g\rangle $\textrm{\ to }$|-\rangle,$ which reduces the
probability amplitude of $|-\rangle $ from the maximum value, instead of resulting in non-negligible probability amplitude in the unwanted $|+\rangle $ state.

\noindent\textbf{Acknowledgements}

\noindent This work is partially supported by MOST (Grants No. 2011CB922104 and No.
2011CBA00200), NSFC (11474154,11474153,91321310,11274156,91021003), PAPD, BK2012013, a doctoral program (20120091110030) and Dengfeng Project B of Nanjing University. S. H. acknowledges partial support
by NSF grant No. PHY-1314861.

\noindent\textbf{Author contributions}

\noindent These authors contributed equally to this work: Guozhu Sun \& Xueda Wen.

\noindent G.S. and S.H. conceived the experiments; G.S. carried out the measurements and analyzed the data with the help of X.W., M.G., D.Z., Y.Y., S.Z., J.C., P.W. and S.H.; X.W. performed the numerical calculations; G.S. and S.H. wrote the paper.

\noindent\textbf{Additional information}

\noindent Competing financial interests: The authors declare no competing financial interests.

\noindent\textbf{Correspondence} Correspondence and requests for materials
should be addressed to G.S. (email: gzsun@nju.edu.cn).

\clearpage
\begin{figure}[Fig1.]
\begin{center}
\includegraphics[width=3.5in]{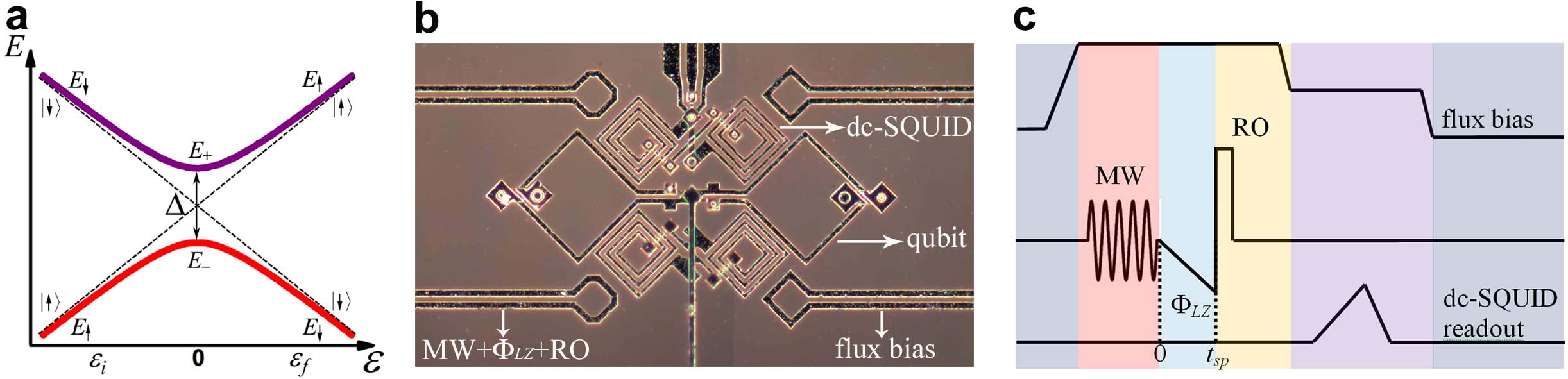}
\caption{\textbf{Circuit and experimental procedure.}
\textbf{(a)} A general avoided energy-level crossing with diabatic basis states
(the dashed lines) and adiabatic basis states (the solid lines). The constant gap $\Delta$ between the two instantaneous eigenenergy states at $\varepsilon =0$ is the tunneling amplitude, i.e., the coupling strength.  \textbf{(b)} Optical micrograph of the sample
with Al/AlO$_{x}$/Al Josephson junctions on the silicon substrate. MW, $\Phi_{LZ}$
and RO represent the microwave pulse to prepare the initial state, the sweeping flux
bias to induce LZT and the $5$-ns flux bias pulse to tilt the potential well in order to readout the qubit state, respectively.
The planar coils and dc-SQUID magnetometer are coupled inductively to the qubit. \textbf{(c)} A typical time
profile of the manipulation and measurement waveforms employed to perform
the single-passage Landau-Zener experiment.}
\end{center}
\end{figure}

\begin{figure}[Fig2.]
\begin{center}
\includegraphics[width=3.5in]{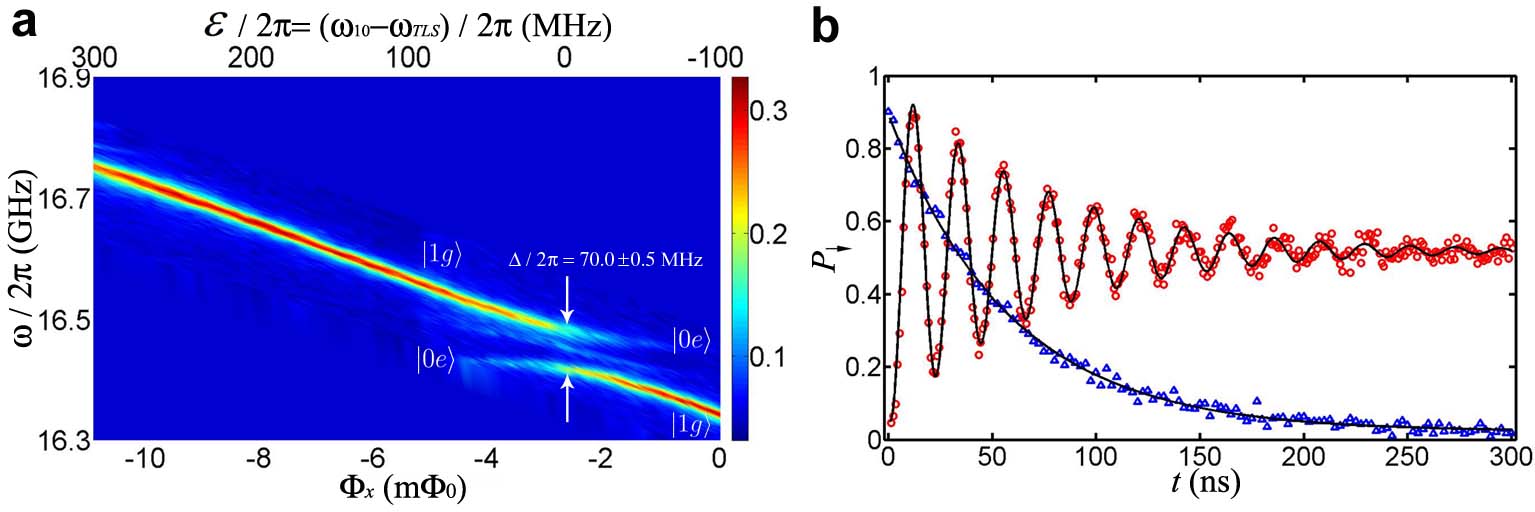}
\caption{\textbf{Spectroscopy and Rabi oscillation.}  \textbf{(a)} Microwave spectroscopy measurement of the coupled qubit-TLS system. The splitting is $\Delta/2\pi=70.0\pm0.5$ MHz centered at
$\omega/2\pi=16.450\pm0.002$ GHz. The beginning point of the flux bias for the single passage LZ sweeping is denoted as $0$ m$\Phi_{0}$, corresponding to $\varepsilon_{i}/2\pi=
(\omega_{10}-\omega_{TLS})/2\pi\approx -100$ MHz in the upper abscissa. \textbf{(b)} Rabi
oscillation and $T_{1}$ at $\varepsilon_{i}$, respectively. The experimental data (the red circles and blue triangles) agree well with the theoretical fits (the black solid lines).}
\end{center}
\end{figure}

\begin{figure}[Fig3.]
\begin{center}
\includegraphics[width=3.5in]{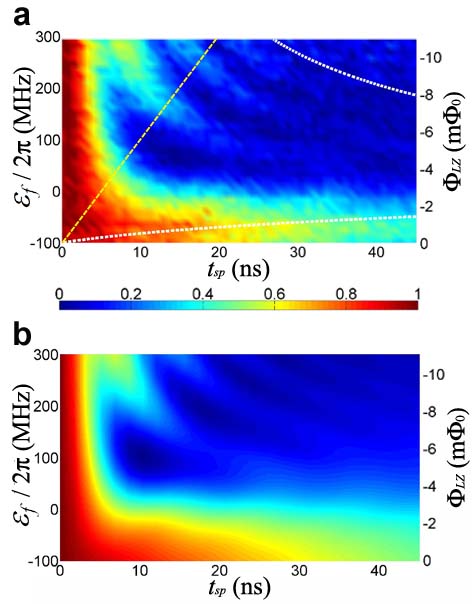}
\caption{\textbf{Coherent LZO.} \textbf{(a)} Experimentally measured $%
P_{\downarrow }$ vs. $\varepsilon _{f}$ and $t_{sp}$. \textbf{(b)} Numerically calculated $P_{\downarrow }$ vs. $\varepsilon _{f}$ and $t_{sp}$ with all
input parameters obtained from the experiment. The white dashed lines
correspond to the value of modified adiabaticity parameter $\alpha' = 10$. Notice
that in the region below (above) the lower (upper) white dashed line, one has $\alpha' >> 1$ thus the system evolves adiabatically and no oscillation in $P_{\downarrow }$ is expected as confirmed
experimentally. The LZ speed $v$ equals the slope of any straight lines
originated from the lower-left corner of the $t_{sp}$-$\varepsilon _{f}$
plane. For example, the yellow dashed line in (a) has $v=400/19.5\approx 20.5
$ MHz/ns. For the sake of clarity, the temporal evolution of the system along the yellow dashed line (constant $v$) is presented separately in Fig. 4(b).}
\end{center}
\end{figure}

\begin{figure}[Fig4.]
\begin{center}
\includegraphics[width=3.5in]{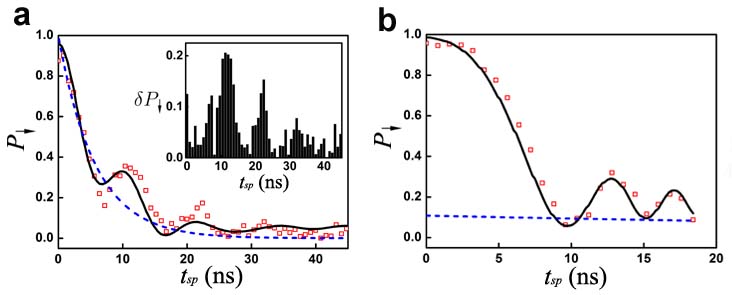}
\caption{\textbf{$P_{\downarrow }$ oscillation at constant LZ speed.} \textbf{(a)} Measured $P_{\downarrow }$ as a function of $t_{sp}$ (the red squares) with $%
\varepsilon _{f}/2\pi = 200$ MHz which clearly shows oscillation in the
region of $\alpha <1$ which compares well with the result of numerical
calculation (the solid line). This is in stark contrast to the smooth
exponential decay expected from the LZ formula (the dashed line). \ The
inset is the difference $\delta P_{\downarrow }=P_{\downarrow }-P_{LZ}$. \textbf{(b)} The measured (the red squares) and numerically calculated (the solid line)
$P_{\downarrow }$ vs. $t_{sp}$ with constant LZ speed $v \approx 20.5$ MHz/ns
corresponding to evolving along the yellow dashed line in Fig. 3(a). \
Again, $P_{\downarrow }$ oscillates in the region where the adiabatic
condition $\alpha >1$ is not satisfied which is not expected from the LZ
formula (the dashed line).}
\end{center}
\end{figure}

\begin{figure}[Fig5.]
\begin{center}
\includegraphics[width=3.5in]{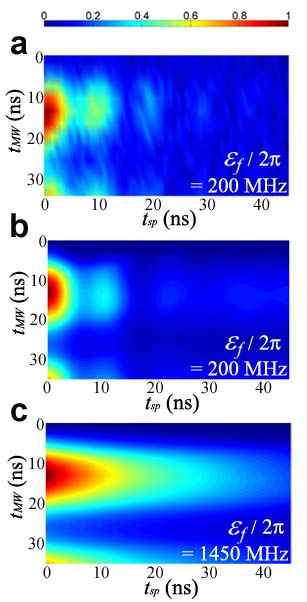}
\caption{\textbf{The effect of $t_{MW}$ on finite time LZT} \textbf{(a)} Experimentally measured and \textbf{(b)} numerically calculated $P_{\downarrow }$ vs.
$t_{MW}$ and $t_{sp}$ for $\varepsilon _{f}/2\pi = 200$ MHz showing
that the effect of imprecise $\pi $ pulse is to reduce the visibility of the
oscillation of $P_{\downarrow }$ vs. $t_{sp}$ by reducing the probability
amplitude of the desired $|-\rangle $ state. Because the microwave pulse is
resonant with the $|0g\rangle \leftrightarrow |-\rangle $ transition while
largely detuned from the $|0g\rangle \leftrightarrow |+\rangle $
transition even a significant deviation from a $\pi $ pulse would only
result in negligible transfer of population to $|+\rangle $. Furthermore,
since $|0g\rangle $ does not participate in the single passage LZ process,
the observed oscillation could neither be due to LZS interference nor
non-negligible population in $|+\rangle $ at the beginning of each $%
\varepsilon $ sweep. For comparison, we also present the numerically
calculated $P_{\downarrow }(t_{MW},t_{sp})$ for $\varepsilon _{f}/2\pi = 1450\approx 20.7\Delta $ in \textbf{(c)}.
The result shows the exponential
decay behavior described by the asymptotic LZ formula as expected for $%
\varepsilon _{f}/\Delta \gg 1.$}
\end{center}
\end{figure}

\end{document}